\documentclass[10pt, conference]{IEEEtran}
\IEEEoverridecommandlockouts

\usepackage{cite}
\usepackage[utf8]{inputenc}
\usepackage[greek,russian,spanish,english]{babel}
\usepackage[OT2,T1, T2A,OT1]{fontenc}
\usepackage{amsmath,amssymb,amsfonts}
\usepackage{algorithmic}
\usepackage{graphicx}
\usepackage{textcomp}
\usepackage{xcolor}
\usepackage{subfig}
\usepackage{float}
\makeatletter
\g@addto@macro\normalsize{%
	\setlength\abovedisplayskip{4pt plus 1pt minus 1pt}%
	\setlength\belowdisplayskip{4pt plus 1pt minus 1pt}%
	\setlength\abovedisplayshortskip{2pt plus 1pt minus 1pt}%
	\setlength\belowdisplayshortskip{2pt plus 1pt minus 1pt}%
	\setlength\jot{1.2pt}
}
\makeatother
\DeclareSubrefFormat{myparens}{#1(#2)}
\def\BibTeX{{\rm B\kern-.05em{\sc i\kern-.025em b}\kern-.08em
    T\kern-.1667em\lower.7ex\hbox{E}\kern-.125emX}}
\begin{document}
{
\title{CKM Beyond Channel Gain: Spatial Correlation Map Construction with Deep Learning}
}

\author{
  \IEEEauthorblockN{
       Zhitong Chen\textsuperscript{1}, 
       Shen Fu\textsuperscript{1}, 
       Yong Zeng\textsuperscript{1,2},
       Xiaoli Xu\textsuperscript{1}, 
       Zhiqiang Wei\textsuperscript{3}
   }

   \IEEEauthorblockA{\textsuperscript{1}National Mobile Communications Research Laboratory, Southeast University, Nanjing 210096, China\\}
	\IEEEauthorblockA{\textsuperscript{2}Purple Mountain Laboratories, Nanjing 211111, China}
  \IEEEauthorblockA{\textsuperscript{3}Xi’an Jiaotong University, Xi'an 710049, China}
  
  Emails: \{220250897; sfu; yong\_zeng; xiaolixu\}@seu.edu.cn, zhiqiang.wei@xjtu.edu.cn}  

\maketitle

\begin{abstract}
Channel knowledge map (CKM) is a promising technique to achieve environment-aware wireless communication and sensing. Constructing the complete CKM based on channel knowledge observations at sparse locations is a fundamental problem for CKM-enabled wireless networks. However, most existing works on CKM construction only consider the special type of CKM, i.e., the channel gain map (CGM), which only records the channel gain value for each location. In this paper, we consider the channel spatial correlation map (SCM) construction, which signifies the location-specific spatial correlation matrix for multi-antenna systems. Unlike CGM construction, constructing SCM poses significant challenges due to its extremely high‐dimensional structure.
To address this issue, we first decompose the high-dimensional SCM into lower-dimensional path gain map (PGM) and path angle map (PAM). Then we propose a deep learning model termed E-SRResNet for constructing high-quality SCM from sparse samples,
which incorporates multi-head attention (MHA) mechanisms and multi-scale feature fusion (MSFF) to accurately model both local and global spatial relationships of channel parameters and complex nonlinear mappings. 
Furthermore, we preprocess the dataset to provide priors including line-of-sight (LoS) map, binary building map and base station (BS) map for the model to reconstruct SCM more accurately.
Simulations conducted on the CKMImageNet dataset demonstrate that the proposed E-SRResNet achieves significant performance improvements over baseline methods.  
Moreover, the cosine similarity between the constructed SCM and the ground truth exceeds 0.8 in most regions, validating the effectiveness of the proposed construction method. 
\end{abstract}


\section{Introduction}

As a foundational statistical metric for wireless communication, the channel spatial correlation matrix describes spatial correlations between antennas in multi-antenna systems, which is a crucial knowledge for beamforming, channel estimation, and resource allocation.
Acquiring channel spatial correlation matrix can substantially reduce the overhead of instantaneous channel state information (CSI) estimation, making it crucial for the sixth generation (6G) mobile communications systems. However, traditional channel spatial correlation matrix estimation algorithms typically require massive real-time channel measurements for the second-order statistical processing, which fails to meet efficiency and latency requirements.
The recently proposed concept of channel knowledge map (CKM) offers a novel paradigm for channel spatial correlation matrix acquisition \cite{9373011}, \cite{10430216}.
CKM constructs a channel knowledge database that characterizes intrinsic wireless propagation properties within specific regions, which is achieved by integrating massive historical channel measurements from all terminals in the target regions with artificial intelligence (AI), big data analytics, and wireless sensing technologies. With CKM, mobile terminals can directly obtain the prior information through physical or virtual location indicators, eliminating redundant real-time environmental sensing or channel estimation procedures, thereby significantly enhancing the efficiency and performance of communication, sensing, localization, and navigation. The spatial correlation
map (SCM) serves as a manifestation of CKM, which tries to learn the channel spatial correlation matrix at each location.


For CKM construction, its methodologies are primarily categorized into model-driven and data-driven approaches. Model-driven approaches such as those in \cite{10530520} and \cite{9771802}, employ analytical models or expectation maximization (EM) algorithms to reconstruct the complete CKM from sparsely observed data. However, these model-based approaches cannot accurately reflect the actual channels in complex environments due to model bias caused by simplified assumptions. 
On the other hand, data-driven methods leverage historical measurements through interpolation or deep learning for map completion. Classical interpolation techniques like k-nearest neighbors (KNN) \cite{Peterson}, kriging \cite{7542153}, and inverse distance weighting  (IDW) \cite{LU20081044} can only fill missing values through spatial smoothing without capturing abrupt or non-stationary channel features. 
In the computer vision community, a class of AI-based methods has been widely used for CKM construction tasks \cite{9972459,9354041,fu2025ckmdiffgenerativediffusionmodel,wang2024deeplearningbasedckmconstruction,10130091,li2025rmtransformeraccurateradiomap,Jin_2025,11141746,10829758}. For instance, the authors in\cite{9972459} introduced a deep convolutional neural network (CNN)-based single-image super-resolution method for end-to-end low resolution (LR)-to-high resolution (HR) mapping, laying groundwork for visual CKM construction. RadioUNet was proposed in \cite{9354041} to efficiently generate channel gain maps (CGMs) in complex urban environments by leveraging end-to-end learning, with city maps, base station locations, and sparse path loss measurements as inputs. The authors in \cite{fu2025ckmdiffgenerativediffusionmodel} proposed CKMDiff, a generative diffusion model for CKM construction by solving inverse problem with learned priors. However, these works mainly focus on the construction of CGM. 

In contrast to the construction for the single dimensional channel knowledge like CGM, constructing SCM poses challenges of increased complexity and prohibitive overhead because the dimension of channel spatial correlation matrix scales quadratically with the number of antennas. Relevant approaches include inferring downlink channel spatial correlation matrix from uplink measurements in frequency division duplex (FDD) systems \cite{8334183} and predicting millimeter wave channel spatial correlation matrix by learning mappings from sub-6GHz data \cite{9420006}, but these approaches are intra-device, i.e.,  they need to predict the SCM within the same device.
Generally, efficient construction and storage of SCM from sparse data remain critical challenges.

In this paper, we consider the SCM completion problem, which is formulated as an image super-resolution task and we predict the SCM by training a deep learning network. We first model the channel spatial correlation matrix in multi-antenna systems based on physical characteristics, transforming it into the tasks of constructing path gain map (PGM) and path angle map (PAM) from sparse samples for those dominant paths. Then we draw an analogy between sparse uniform sampling and image super-resolution, and propose an  architecture termed E-SRResNet tailored to wireless system, integrating multi-head attention (MHA) and multi-scale feature fusion (MSFF) to simultaneously model local and global spatial relationships.
Utilizing the CKMImageNet dataset \cite{11184538}, which provides comprehensive CKMs including PGMs, PAMs and time of arrival (ToA) maps, we extract PGM and PAM for primary and secondary paths, and generate the line-of-sight (LoS) map, the binary building map, and the base station (BS) map to provide priors for the model. Numerical results demonstrate superior performance of our model over the baseline methods, validating its effectiveness in high-accurate sparse SCM completion.

\section{System model}

\subsection{Channel Spatial Correlation Matrix Model}\label{AA}
\begin{figure}[htbp]
	\centering
	\includegraphics[scale=0.25]{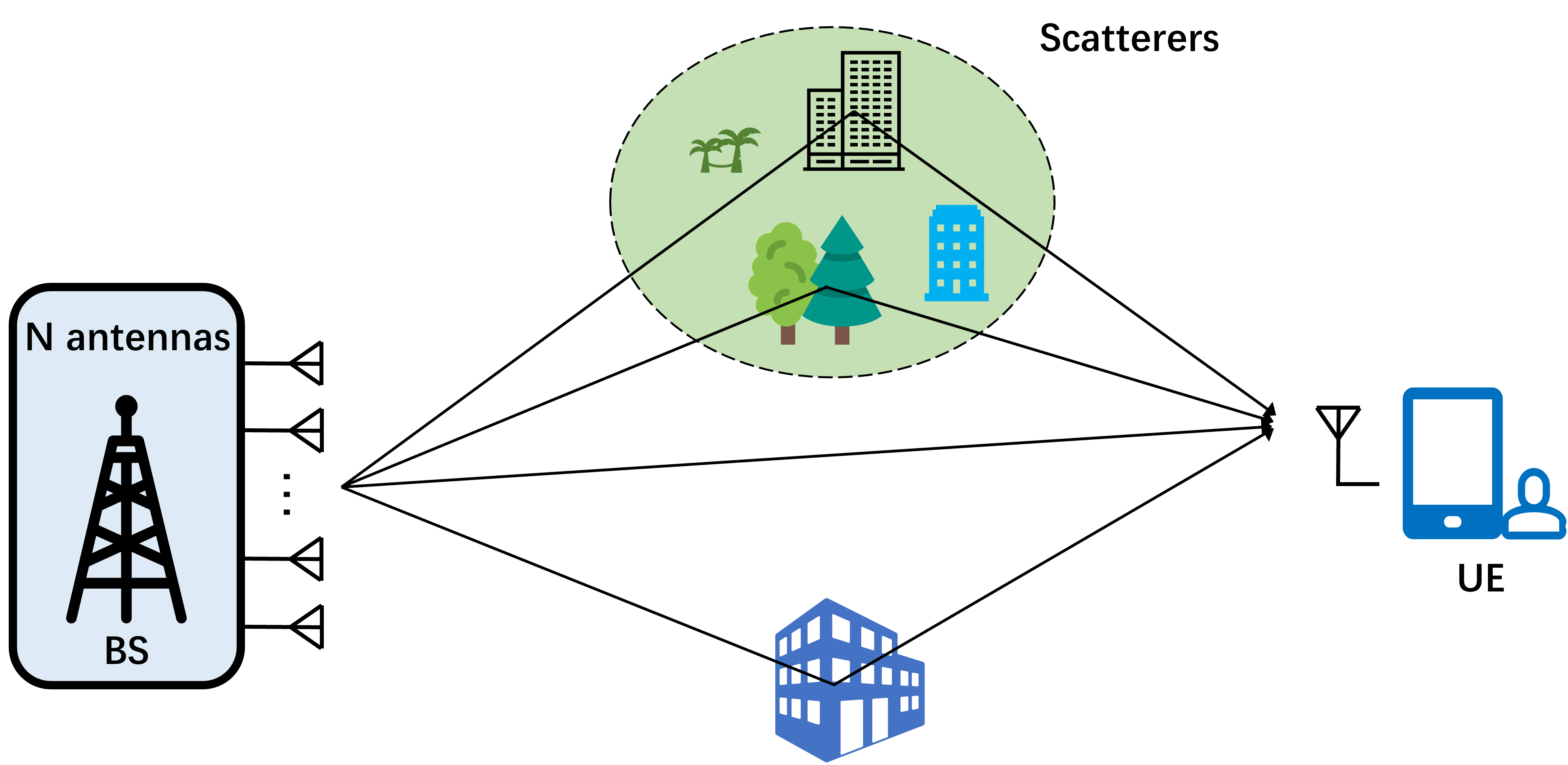}
	\caption{Illustration of the scenario setup.}
	\label{CCM scen}
\end{figure}
Consider massive MIMO scenario, where the BS is equipped with $N\gg 1$ antennas, as shown in Fig.~\ref{CCM scen}. 
Let $\theta_\ell$ and $\alpha_\ell$ denote the angle of arrival (AoA) and complex gain of the $\ell$-th path, respectively. The channel vector $\mathbf{h}$ can be expressed as
\begin{equation}
	\mathbf{h} = \sum_{\ell=0}^{L-1}\alpha_\ell\mathbf{a}(\theta_\ell) =  \sum_{\ell=0}^{L-1}\left|\alpha_\ell\right|\mathbf{a}(\theta_\ell)e^{j\phi_\ell},\label{h_model}
\end{equation}
where $L$ denotes the number of paths, $\phi_\ell$ denotes the random phase of the $\ell$-th path, and $\mathbf{a}(\theta_\ell)$ is the steering vector at the BS.
For uniform linear array (ULA) deployed at the BS, the array steering vector of $\ell$-th path is defined as
\begin{equation}
	\mathbf{a}(\theta_\ell) = \left[ 1, \; e^{\,j\frac{2\pi d}{\lambda}\cos\theta_\ell}, \;  \dots, \; e^{\,j\frac{2\pi d}{\lambda}(N-1)\cos{\theta_\ell}} \right]^T,\label{ULA}
\end{equation}
where $d$ is the array elements spacing and $\lambda$ is the carrier wavelength. Channel spatial correlation matrix is defined as
\begin{equation}
	\mathbf{R} = \mathbb{E}\left[ \mathbf{h} \mathbf{h}^H \right].
\end{equation}

To learn and compute such a high-dimensional matrix directly is difficult. Therefore, under the assumption of multi-path sparsity, channel spatial correlation matrix is given by
\begin{equation}
	\mathbf{R} = \mathbb{E}\left[ \sum_{\ell,k}^{} \left|\alpha_\ell\right|\left|\alpha_k\right| \mathbf{a}(\theta_\ell)\mathbf{a}^H(\theta_k) e^{j(\phi_\ell - \phi_k)} \right].
\end{equation}

To handle the cross-terms in the equation above, we make the following assumptions:
\begin{itemize}
	\item Amplitude-phase independence: 
	$\phi_\ell \perp\!\!\!\perp \left|\alpha_\ell\right|$
	\item Uniformly distributed phase:
	$\phi_\ell \sim \mathcal{U}[0, 2\pi)$	
	\item Phase independence:
	$\mathbb{E}[e^{j(\phi_\ell - \phi_k)}] = \delta(\ell-k)$
\end{itemize}

Under the above assumptions, the channel spatial correlation matrix simplifies to
\begin{equation}
	\mathbf{R} = \sum_{\ell=0}^{L-1} \mathbb{E}\left[|\alpha_\ell|^2\right] \mathbf{a}(\theta_\ell)\mathbf{a}^H(\theta_\ell). 
	\label{Rhh_f}
\end{equation}
The $(m,n)$-th element of $\mathbf{R}$ is given by
\begin{equation}
	[\mathbf{R}]_{m,n} = \sum_{\ell=0}^{L-1} \mathbb{E}\left[|\alpha_\ell|^2\right] e^{\,j\frac{2\pi d}{\lambda}(m-n)\cos\theta_\ell}.\label{Rhh_ula}
\end{equation}

Now, the completion of SCM is recast as PGM and PAM completion problem for $L$ dominant paths where $L\ll N$.
This modeling approach decomposes the channel spatial correlation matrix into physically meaningful parameters, offering two key advantages over alternative methods like eigen value decomposition (EVD), which include
\begin{itemize}
	\item Physical interpretability: Each term $\mathbf{a}(\theta_\ell)\mathbf{a}^H(\theta_\ell)$ directly corresponds to a physical propagation path.
	\item Structure preservation: It can preserves array geometry properties (e.g., Hermitian structure for ULA)
\end{itemize}
\subsection{Channel Correlation Matrix Completion Model}
\begin{figure}[!t]
	\centering
	\includegraphics[scale=0.6]{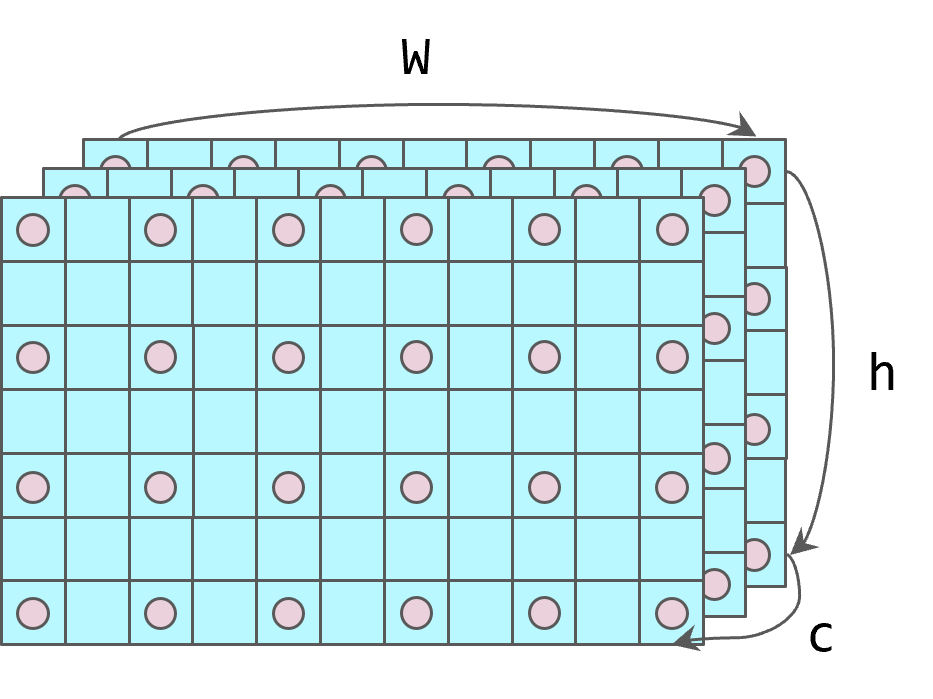}
	\caption{Illustration of uniform sparse sampling grid of CKM.}
	\label{sparse sample}
\end{figure}

\begin{figure}[!t]
	\centering
	\includegraphics[scale=0.6]{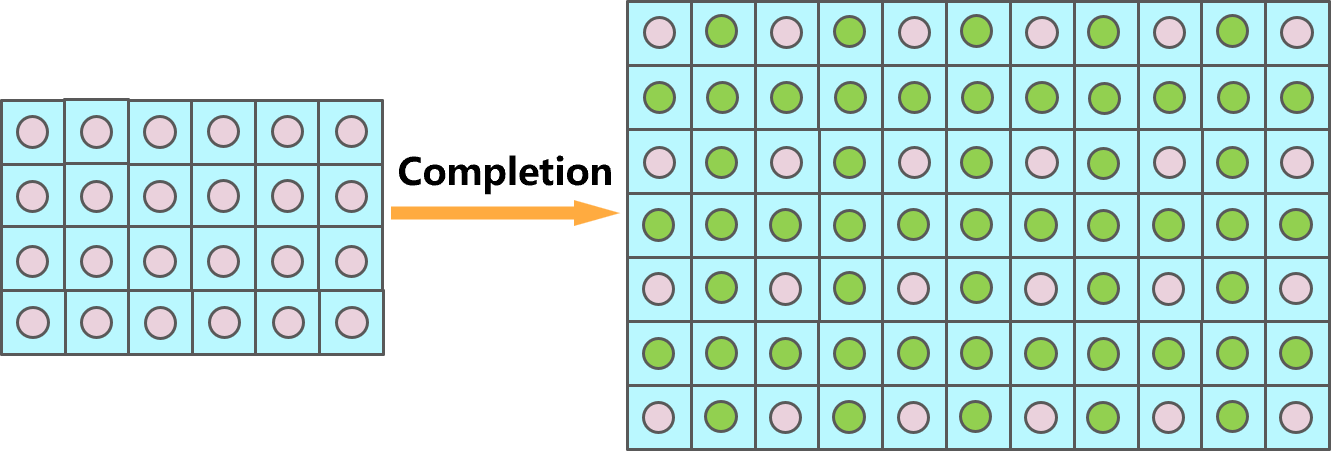}
	\caption{Illustration of SCM completion.}
	\label{CCM Comp}
\end{figure}

We consider to construct CKM from sparse samples within a given region. The continuous geographical region covered by the BS is discretized into uniform 2D grids. The desired complete CKM should cover grids with spatial resolution $w\times h$. Each grid corresponds to a $c$-dimensional channel knowledge such as path gain and AoA. The complete CKM tensor of the region can be defined as $\mathcal{X}\in \mathbb{R}^{w\times h\times c}$.
Sparse sampling is typically performed within the region and yield sparse data $\mathcal{Y}\in \mathbb{R}^{w'\times h'\times c}$, where $w'< w$ and $h'< h$. Fig.~\ref{sparse sample} illustrates a uniform sparse sampling scheme, where red circles indicate sampled locations. Therefore, the sampling process is expressed by
\begin{equation}
\mathcal{Y}_{k} = \mathcal{X}[i_k, j_k, :], \quad \forall k \in \{1,2,\dots,w'\times h'\},
\end{equation}
where $(i_k,j_k)$ denotes the $k$-th sampled location and $\mathcal{Y}_k$ denotes the $k$-th sampled data.
Based on the above CKM sparse sampling model and channel spatial correlation matrix model, the task of SCM completion can be modeled as a tensor completion task as shown in Fig.~\ref{CCM Comp}. 
Specifically, in the $w\times h$ uniform grids,  $\mathbf{q}_{i,j}$ denotes the coordinate of uniformly observed grid point and the channel spatial correlation matrix value at each location is $\mathcal{M}_{i,j,:,:}$, where $\mathcal{M}_{i,j,:,:} = f(\mathbf{q}_{i,j})$. $f(\cdot)$ is a function mapping the coordinate to channel spatial correlation matrix. 

The SCM completion problem aims to reconstruct a complete four-dimensional tensor $\mathcal{H} \in \mathbb{C}^{w \times h \times c \times L}$ which matches the values at all observed locations  $(i,j) \in \Omega$, where $\Omega \subseteq \{1,\dots,w\}\times\{1,\dots,h\}$ is the index set of observed locations, which is given by

\begin{equation}
	\mathcal{H}_{i,j,:,:} = \mathcal{M}_{i,j,:,:}, \quad \forall (i,j) \in \Omega.
\end{equation}


\section{Spatial Correlation Map Completion Based on E-SRResNet}
\subsection{Network Architecture}

\begin{figure*}[htbp]
	\centering
	\includegraphics[scale=0.7]{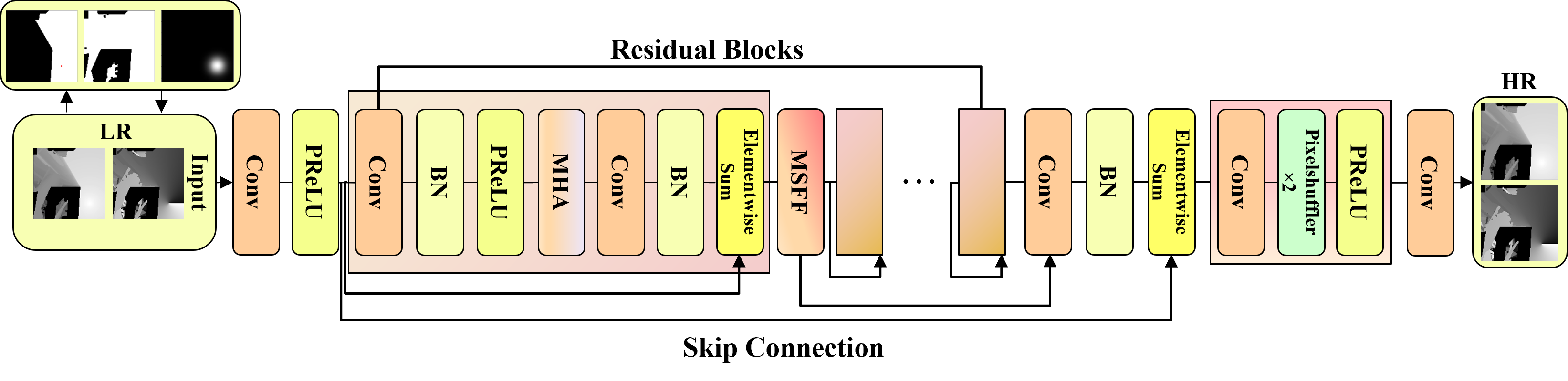}
	\caption{Architecture of E-SRResNet.}
	\label{Architecture}
\end{figure*}
Due to the inherent physical properties of electromagnetic wave propagation, PGM and PAM exhibit significant long-range spatial dependencies and multi-scale features. In order to accurately construct SCM from the uniformly observed data, we propose a neural network architecture termed E-SRResNet based on SRResNet \cite{ledig2017photorealisticsingleimagesuperresolution} by introducing MHA modules and MSFF modules specifically for communication scenarios. 
The model input of the primary path is a five-channel map: PGM, PAM, LoS map, binary building map and BS map, while the model input for the secondary path is  a four-channel map: PGM, PAM, binary building map and BS map. The output is a two-channel map composed of the completed PGM and PAM of each path and then generate SCM. The MHA and MSFF modules enable the network to capture both long-range context and multi-scale structure necessary for high-fidelity SCM reconstruction. The overall architecture of E-SRResNet is shown in Fig.~\ref{Architecture}
\subsection{Multi-head Attention And Multi-scale Feature Fusion}
The MHA module is inserted into each residual block, with the number of attention heads set to 4. The self-attention mechanism computes correlations between positions in feature maps, enabling the network to focus on global information of the input feature map when generating each output point \cite{NIPS2017_3f5ee243}. Due to long-range channel characteristic correlations caused by building obstructions or multipath propagation effects in physical environments, PGM and PAM exhibit complex spatial dependencies at distant locations. 


To extract the features of different scales \cite{9551165}, MSFF is introduced to process large-scale LoS patterns in open spaces where signals propagate directly with smooth path gain variations, alongside fine-grained non line-of-sight (NLoS) details around buildings where multiple reflections create complex angle discontinuities and interference patterns. Three such modules are inserted after the 4th, 8th, and 12th residual blocks in the 16-block backbone to capture features at different processing stages. 
Each module contains convolutional branches with kernel sizes of 1, 3, and 3 respectively, where the two 3×3 convolutional kernels have dilation rates of 2 and 5.
This dilation pyramid achieves 7×7 to 15×15 receptive fields through sparse sampling, expanding coverage without parameter inflation. Through channel concatenation fusion, local and global features are fully fused, maximizing the utilization of low-resolution images for finer super-resolution reconstruction.


\section{Spatial Correlation Map Dataset}
\begin{figure}[!t]
	\centering
	\subfloat[PGM$^{1}$]{
		\begin{minipage}{0.35\linewidth}
			\centering
			\includegraphics[width=0.9\linewidth]{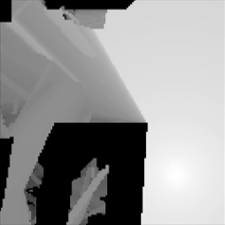}
			\label{PGM1}
		\end{minipage}
	}
	\subfloat[PAM$^{1}$]{
		\begin{minipage}{0.35\linewidth}
			\centering
			\includegraphics[width=0.9\linewidth]{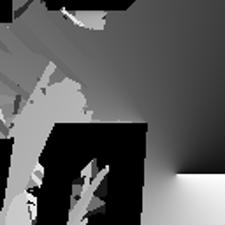}
			\label{PAM1}
		\end{minipage}
	}
	\vspace{-0.9em} 
	\subfloat[PGM$^{2}$]{
		\begin{minipage}{0.35\linewidth}
			\centering
			\includegraphics[width=0.9\linewidth]{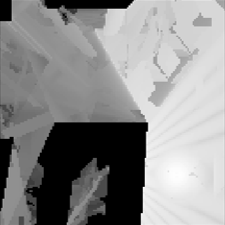}
			\label{PGM2}
		\end{minipage}
	}
	\subfloat[PAM$^{2}$]{
		\begin{minipage}{0.35\linewidth}
			\centering
			\includegraphics[width=0.9\linewidth]{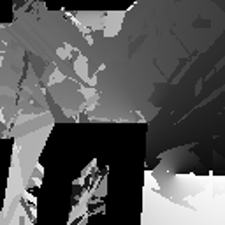}
			\label{PAM2}
	\end{minipage}}
	\caption{Visualization of dataset.}
	\label{dataset}
\end{figure}

The dataset employed is CKMImageNet \cite{11184538}, a location-based CKM dataset constructed using the commercial ray-tracing software Wireless Insite, which comprises channel gain maps, AoA and AoD information, as well as scene views and parameter configurations. In this dataset, gain values range from -250 dB to -50 dB and are linearly mapped to the pixel range $[0,1]$. Angle values are handled by assigning building-region pixels a value of -200°, and linearly mapped the angular interval from -200° to 180° to the pixel range $[0,1]$.
To complete SCM , we read the path gain data, path angle data, user equipment (UE) height data, and BS height data from the dataset.
One representative pair of PGM and PAM for the primary and secondary paths is shown in Fig.~\ref{dataset}.

\begin{figure}[H]
	\centering
	\subfloat[LoS map]{
		\begin{minipage}{0.3\linewidth}
			\centering
			\includegraphics[width=0.9\linewidth]{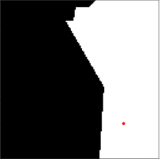}
			\label{lomm}
		\end{minipage}
	}
	\subfloat[Binary building map]{
		\begin{minipage}{0.3\linewidth}
			\centering
			\includegraphics[width=0.9\linewidth]{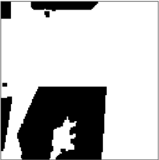}
			\label{bim}
		\end{minipage}
	}
	\subfloat[BS map ($\sigma^2 = 5$)]{
		\begin{minipage}{0.3\linewidth}
			\centering
			\includegraphics[width=0.9\linewidth]{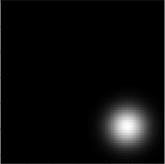}
			\label{bsm}
		\end{minipage}
	}
	\caption{Illustration of prior map.}
	\label{prior}
\end{figure}

The signal propagation in LoS regions follows direct paths with simple characteristics, while in NLoS regions exhibits complex scattering and reflection phenomena. The map of LoS labels can effectively distinguish between LoS and NLoS regions and will offer extra information to enhance the network's ability to learn high-frequency boundary information at the LoS and NLoS transition zone and region-specific image features under different propagation mechanisms.
To construct the LoS map, the pixel with maximum brightness in the PGM image is first identified, corresponding to the BS location where path gain is the highest. According to UE and BS height data, the free-space gain is calculated using the Friis formula.
By comparing the calculated gain with the strongest path gain, we determine whether the dominant path is LoS. Pixels are labeled as 1 for LoS paths which are white pixels and 0 for NLoS paths which are black pixels, generating the LoS map, with the calculated BS position marked by a red pixel as shown in Fig.~\ref{lomm}.
Buildings induce shadowing and multipath effects while exhibiting pixel-level discontinuities at boundaries, complicating perimeter feature learning. Incorporating the binary building map enables the network to capture reflection patterns and enhance high-frequency edge feature extraction. To generate the binary building map, a fixed threshold is applied to the PGM for binarization, partitioning image pixels into white and black classes, where pixel values of building regions are set to 0 and others are set to 1. After thresholding, the binary building map is obtained as shown in Fig.~\ref{bim}.

Since both PGM and PAM characterize signal propagation centered at the BS, incorporating BS location as input enables the network to learn distance-dependent features such as path loss. To achieve this, we generate BS maps through Gaussian encoding, which can prevent the loss of BS location information during downsampling. The encoding procedure places a two‐dimensional Gaussian centered at the BS location which is brightest in the grayscale map and assigns values to all pixels in the image with a variance of $\sigma^2$ as shown in Fig.~\ref*{bsm}.
Finally, for the primary path, the five-channel image is formed by concatenating the PGM, PAM, LoS map, binary building map, and BS map.
Additionally, for the secondary path and beyond, the four-channel image is formed by concatenating the PGM, PAM, binary building map and BS map, as there is no LoS component in these paths.

\section{Numerical Results}
\subsection{Setup}
\begin{figure}[!t]
	\centering
	\subfloat[]{
		\begin{minipage}{0.3\linewidth}
			\centering
			\includegraphics[width=0.9\linewidth]{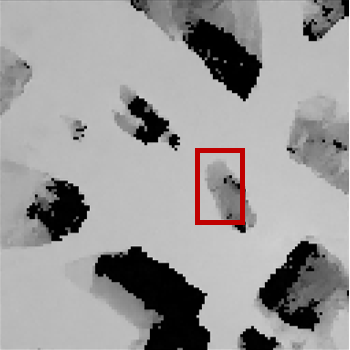}
			\label{nobld}
		\end{minipage}
	}
	\subfloat[]{
		\begin{minipage}{0.3\linewidth}
			\centering
			\includegraphics[width=0.9\linewidth]{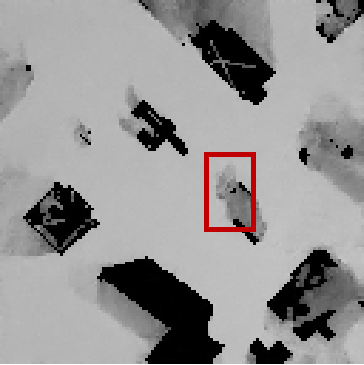}
			\label{bld2}
		\end{minipage}
	}
	\subfloat[]{
		\begin{minipage}{0.3\linewidth}
			\centering
			\includegraphics[width=0.9\linewidth]{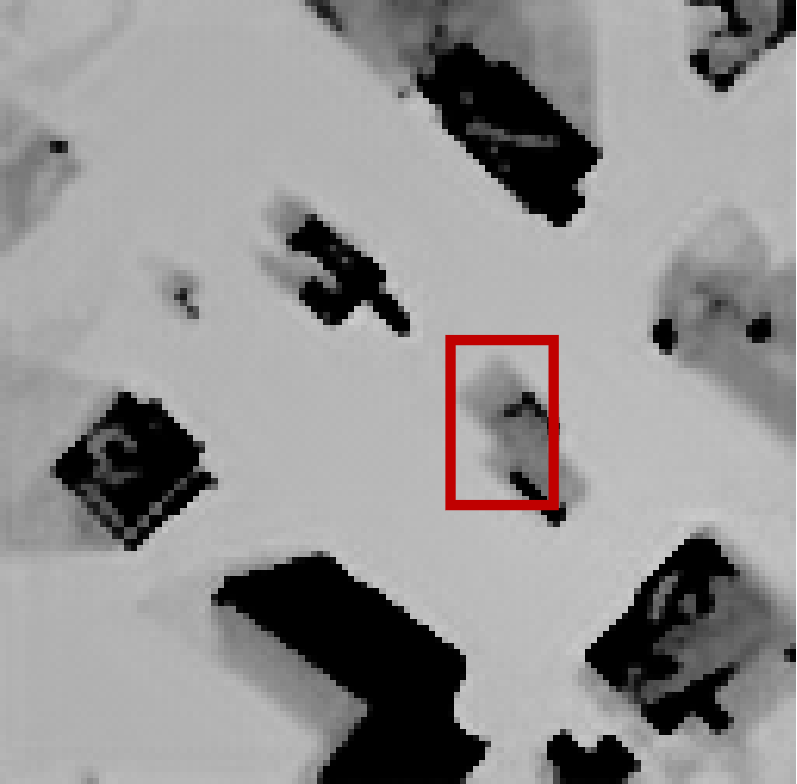}
			\label{bld_los}
		\end{minipage}
	}
	\caption{Ablation study results of input priors: (a) baseline input (PGM and PAM); (b) baseline with binary building map; (c) baseline with binary building map and LoS map.}
	\label{ablation}
\end{figure}

For 2× super-resolution training, we utilize over 22,000 preprocessed images from the CKMImageNet dataset at 28GHz, which are randomly split into training and testing sets
with proportions of 90\% and 10\%.
The network takes 5-channel low-resolution input data as $64\times 64$ and generates two-channel PGM and PAM as $128\times 128$ and the variance $\sigma^2$ of the BS map is set to 5.
To evaluate the contribution of each input map, we conduct an ablation study by progressively adding the binary building map and LoS map to the base two-channel input, the visualization of which is shown in Fig.~\ref{ablation}. In order to focus on pixel-level accuracy in SCM completion, the MSE between the ground truth and the super-resolution result is selected as the loss function. 
We train the E-SRResNet through the following two steps. First, to train the model for the primary path, we use 16 residual blocks with 64 channels and batch size set to 32. Using Adam optimizer with initial learning rate of 0.0002 for 200000 iterations. The scheduler is set to reduce the learning rate by half every 5 epochs when the validation loss stops decreasing. Second, to train the model for the secondary path, the number of residual blocks increases to 18 and channels expand to 128 for enhanced extraction of complex features. Other parameters remain consistent with first step. The powers of the third and other paths are much smaller than that of the first path and can therefore be neglected.

\subsection{Results}
\begin{figure*}[htbp]
	\centering
	\subfloat[PGM$^1_\text{E-SRResNet}$]{
	\begin{minipage}{0.20\linewidth}
		\centering
		\includegraphics[width=0.8\linewidth]{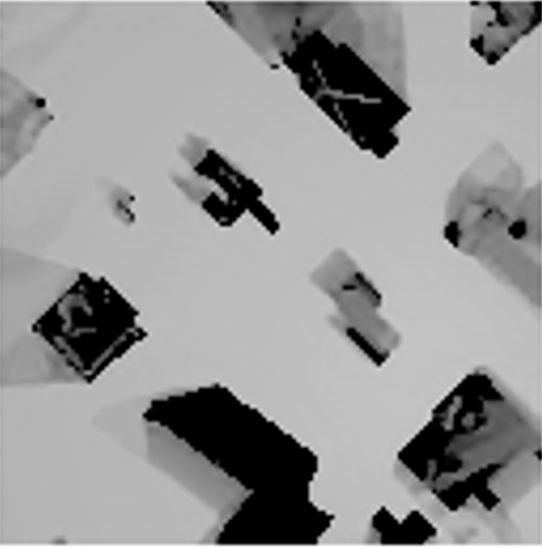}
		\label{PGM1SR}
	\end{minipage}
	}
	\subfloat[PAM$^1_\text{E-SRResNet}$]{
	\begin{minipage}{0.20\linewidth}
		\centering
		\includegraphics[width=0.8\linewidth]{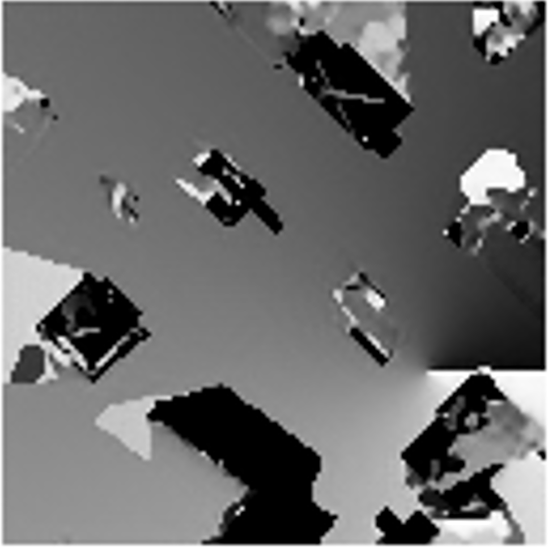}
		\label{PAM1SR}
	\end{minipage}
	}
	\subfloat[PGM$^2_\text{E-SRResNet}$]{
	\begin{minipage}{0.20\linewidth}
		\centering
		\includegraphics[width=0.8\linewidth]{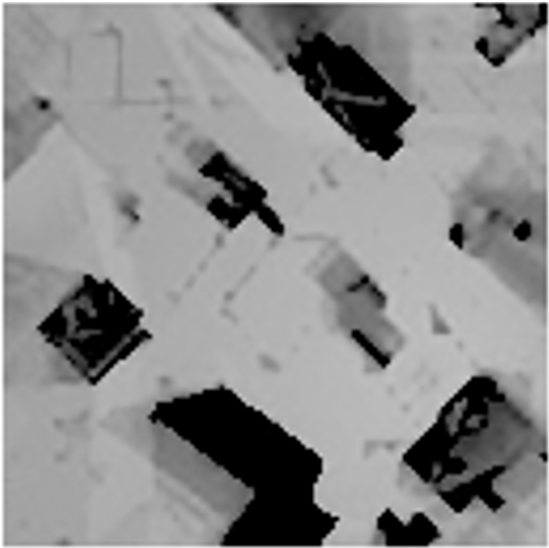}
		\label{PGM2SR}
	\end{minipage}
	}
	\subfloat[PAM$^2_\text{E-SRResNet}$]{
	\begin{minipage}{0.20\linewidth}
		\centering
		\includegraphics[width=0.8\linewidth]{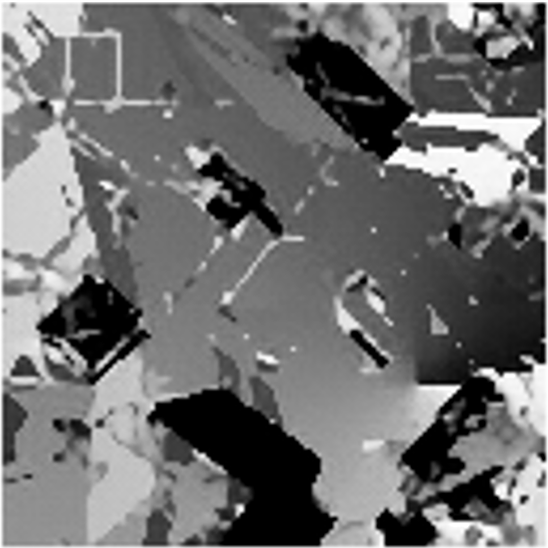}
		\label{PAM2SR}
	\end{minipage}
	}
	\vspace{-0.5em} 
	\qquad
	\subfloat[PGM$^1_\text{bic}$]{
	\begin{minipage}{0.20\linewidth}
		\centering
		\includegraphics[width=0.8\linewidth]{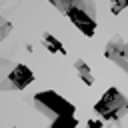}
		\label{Bic_1PGM}
	\end{minipage}
	}
	\subfloat[PAM$^1_\text{bic}$]{
		\begin{minipage}{0.20\linewidth}
			\centering
			\includegraphics[width=0.8\linewidth]{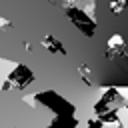}
			\label{Bic_1PAM}
		\end{minipage}
	}
	\subfloat[PGM$^2_\text{bic}$]{
		\begin{minipage}{0.20\linewidth}
			\centering
			\includegraphics[width=0.8\linewidth]{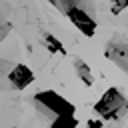}
			\label{Bic_2PGM}
		\end{minipage}
	}
	\subfloat[PAM$^2_\text{bic}$]{
		\begin{minipage}{0.20\linewidth}
			\centering
			\includegraphics[width=0.8\linewidth]{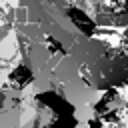}
			\label{Bic_2PAM}
		\end{minipage}
	}
	\vspace{-0.5em} 
		\qquad
	\subfloat[PGM$^1_\text{KNN}$]{
		\begin{minipage}{0.20\linewidth}
			\centering
			\includegraphics[width=0.8\linewidth]{Bic_1PGM.png}
			\label{KNN_1PGM}
		\end{minipage}
	}
	\subfloat[PAM$^1_\text{KNN}$]{
		\begin{minipage}{0.20\linewidth}
			\centering
			\includegraphics[width=0.8\linewidth]{Bic_1PAM.png}
			\label{KNN_1PAM}
		\end{minipage}
	}
	\subfloat[PGM$^2_\text{KNN}$]{
		\begin{minipage}{0.20\linewidth}
			\centering
			\includegraphics[width=0.8\linewidth]{Bic_2PGM.png}
			\label{KNN_2PGM}
		\end{minipage}
	}
	\subfloat[PAM$^2_\text{KNN}$]{
		\begin{minipage}{0.20\linewidth}
			\centering
			\includegraphics[width=0.8\linewidth]{Bic_2PAM.png}
			\label{KNN_2PAM}
		\end{minipage}
	}
	\vspace{-0.5em} 
	\qquad
	\subfloat[PGM$^1_\text{gt}$]{
	\begin{minipage}{0.20\linewidth}
		\centering
		\includegraphics[width=0.8\linewidth]{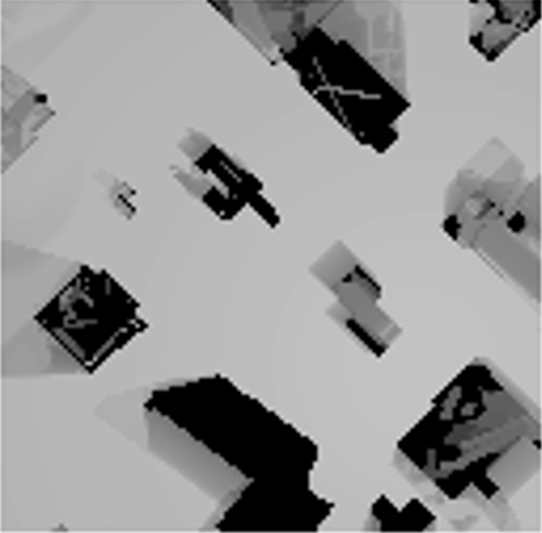}
		\label{PGM1HR}
	\end{minipage}
	}
	\subfloat[PAM$^1_\text{gt}$]{
	\begin{minipage}{0.20\linewidth}
		\centering
		\includegraphics[width=0.8\linewidth]{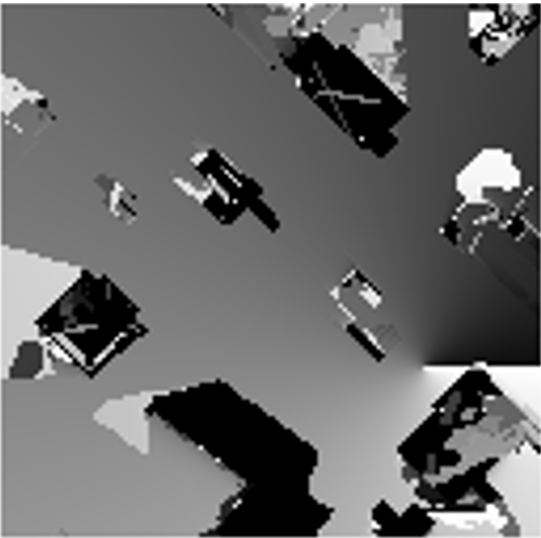}
		\label{PAM1HR}
	\end{minipage}
	}
	\subfloat[PGM$^2_\text{gt}$]{
	\begin{minipage}{0.20\linewidth}
		\centering
		\includegraphics[width=0.8\linewidth]{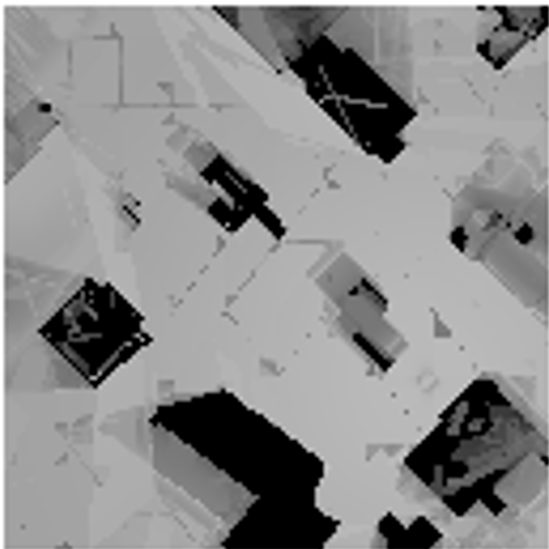}
		\label{PGM2HR}
	\end{minipage}
	}
	\subfloat[PAM$^2_\text{gt}$]{
	\begin{minipage}{0.20\linewidth}
		\centering
		\includegraphics[width=0.8\linewidth]{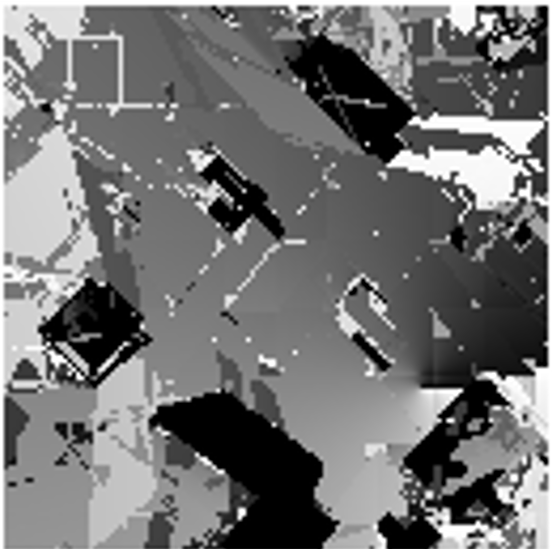}
		\label{PAM2HR}
	\end{minipage}
	}
	\vspace{-0.2em} 
	\caption{Visualization of the PGM and PAM completion results for the primary and secondary paths.}
	\label{result}
\end{figure*}

\vspace{-0.4em}
The inference results of the two path are shown in Fig.~\ref{result}. 
The MSE between the super-resolution results and ground truth for both primary path and secondary path is calculated, and then mapped to actual gain and angle metrics, while excluding error contributions from building structures. 

It can be seen that while the primary path achieves effective completion in most regions, performance degrades significantly around buildings, at NLoS or LoS transition zones, and in select NLoS areas. The secondary path exhibits marginally weaker performance than the primary path, due to the extreme learning complexity caused by intricate scattering, reflection, and diffraction  around buildings. Particularly challenging is PAM characterized by rapid variations that leads to greater learning difficulties than PGM. 

Comparative results among the bicubic interpolation, the KNN and the E-SRResNet proposed in this paper are presented in Table~\ref{tab1}. For the primary path super-resolution task, the E-SRResNet demonstrates significant performance improvement: approximately 41.3\% improvement in PGM completion and 54.3\% improvement in PAM completion compared to the KNN and 23.7\% improvement in PGM completion and 47.7\% improvement in PAM completion compared to the bicubic.

\begin{table}[!t]
	\caption{The completion RMSE of different methods}
	\begin{center}
		\begin{tabular}{|c|ccc|}
			\hline
			\textbf{}& {\textbf{Bicubic}} & \textbf{KNN} & \textbf{E-SRResNet}\\
			\hline 
			\textbf{PGM$^{1}$} & 2.8782 & 3.7431 & \textbf{2.1962} \\
			\hline
			\textbf{PAM$^{1}$} & 18.2459 & 20.8850 & \textbf{9.5341} \\
			\hline
			\textbf{PGM$^{2}$} & 4.3460 & 5.0913 & \textbf{3.6890} \\
			\hline
			\textbf{PAM$^{2}$} & 35.9281 & 39.9711 & \textbf{25.4013} \\
			\hline
		\end{tabular}
		\label{tab1}
	\end{center}
\end{table}
\vspace{-0.4em}

\begin{figure}[htbp]
	\centering
	\includegraphics[scale=0.25]{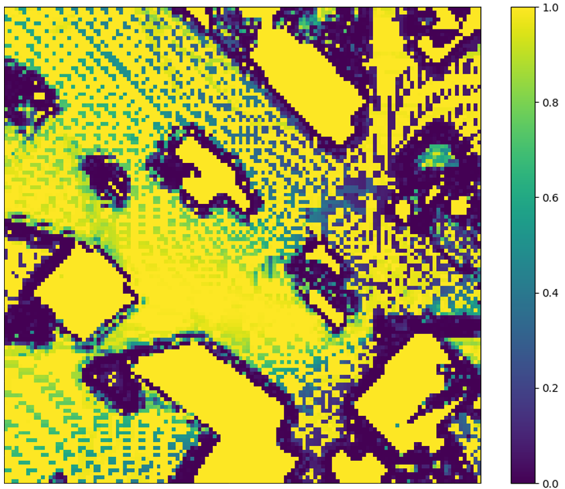}
	\caption{Cosine similarity map.}
	\label{cosine}
\end{figure}

To evaluate the SCM completion performance, we calculate the cosine similarity between the completed SCM results and the ground truth SCM, which is an effective metric for consistency between two matrices. Assuming a ULA deployed at the BS, the SCM can be synthesized according to \eqref{Rhh_ula}.
With 64 antennas at the BS and an inter-element spacing of $d = \lambda /2$, we generate the SCM using the completed PGM and PAM of both primary and secondary paths from the the test sample. Simultaneously, the ground truth SCM is computed. The cosine similarity $\text{Corr}_{i}$ is defined as
\begin{equation}
	\text{Corr}_{i} = \frac{\text{tr}(\widehat{\mathbf{R}}_i \mathbf{R}_i)}{\|\widehat{\mathbf{R}}_i\|_F \|\mathbf{R}_i\|_F},
\end{equation}
where $\text{tr}(\cdot)$ denotes the matrix trace, $\widehat{\mathbf{R}}_i$ represents the completed SCM result at the $i$-th position, $\mathbf{R}_i$ corresponds to the ground truth SCM result at the $i$-th position and $\|\cdot\|_F$ indicates the Frobenius norm of a matrix. The cosine similarity is visualized as Fig.~\ref{cosine}.
The cosine similarity in most regions exceeds 0.8, but it degrades significantly in NLoS regions and near building boundaries, consistent with the PGM and PAM completion results. 

\section{Conclusion}
\vspace{-0.4em}
In this paper,  we tackle the challenging problem of SCM construction by decomposing the complex high-dimensional SCM completion task into sparse completion of PGM and PAM, and propose an E-SRResNet model that incorporates MHA mechanisms and MSFF to achieve SCM completion from sparse samples. Additionally, we generate LoS map, binary building map and BS map as network inputs to improve completion accuracy. Numerical results demonstrate that the E-SRResNet achieves superior performance in the sparse SCM completion task. In the future work, we will extend our study to more sparse sampling and more propagation paths to improve the scalability and accuracy of our model.
\vspace{-0.4em}
%
\section*{Acknowledgment}
\vspace{-0.4em}
This work was supported by the National Natural Science Foundation of China under Grant 62571116, and by the Fundamental Research Funds for the Central Universities under Grants 2242022k60004 and 3204002004A2.
\vspace{-0.4em}
\bibliographystyle{IEEEtran}
\bibliography{IEEEabrv,myref}
\end{document}